\documentclass[prb,preprint]{revtex4} 

\usepackage[utf8]{inputenc}
\usepackage[english]{babel}
\usepackage{amsmath}
\usepackage[caption = false]{subfig}
\usepackage{graphicx,epstopdf}
\usepackage{blindtext}
\usepackage{lipsum}
\usepackage{amsfonts}
\usepackage{bbm}
\usepackage{amssymb}
\usepackage{enumerate}
\usepackage{color}
\usepackage{latexsym}

\newcommand{\Omath}{\mathbbm{O}}

\newcommand{\Mcal}{\mathcal{M}}

\newcommand{\Ocal}{\mathcal{O}}

\newcommand{\Ucal}{\mathcal{U}}

\newcommand{\1}{\mathbbm{1}}

\newcommand{\ket}[1]{| #1 \rangle}
\newcommand{\bra}[1]{\langle #1 |}

\newcommand{\interpro}[2]{\langle #1 | #2 \rangle}
\newcommand{\ketus}[0]{|\!\! \uparrow \rangle}
\newcommand{\ketds}[0]{|\!\! \downarrow \rangle}
\usepackage[normalem]{ulem}

\begin{document}


\title{Simulating single-spin dynamics on an IBM five-qubit chip}

\author{Émerson M. Alves}
\email{emerson.amn@gmail.com} 
\affiliation{Departamento de Física, Universidade Regional do Cariri, Av. Leão Sampaio, 107, Triângulo, 63041-145,Juazeiro do Norte, Ceará, Brazil}

\author{Francisco D. S. Gomes}
\email{coltan1804@gmail.com} 
\affiliation{Departamento de Física, Universidade Regional do Cariri, Av. Leão Sampaio, 107, Triângulo, 63041-145,Juazeiro do Norte, Ceará, Brazil}

\author{Hércules S. Santana}
\email{hercules-santana@hotmail.com} 
\affiliation{Departamento de Física, Universidade Regional do Cariri, Av. Leão Sampaio, 107, Triângulo, 63041-145,Juazeiro do Norte, Ceará, Brazil}

\author{Alan C. Santos}
\email{ac_santos@id.uff.br}
\affiliation{Instituto de F\'{i}sica, Universidade Federal Fluminense, Av. Gal. Milton Tavares de Souza s/n, Gragoat\'{a}, 24210-346 Niter\'{o}i, Rio de Janeiro, Brazil}


\date{\today}

\begin{abstract}
In this paper we show how the IBM superconducting chips can be a powerful tool for teaching foundations of quantum mechanics for undergraduate students (for graduates as well, in some cases). To this end, we briefly discuss about the main elements of the IBM Quantum Experience platform necessary to understand this paper, i.e., how to implement operations and single-qubit measurements. We experimentally study the dynamics of single spin systems interacting with static and time-dependent magnetic fields. First, we study the resonant behavior of a single spin coupled to a time-dependent rotating magnetic field. To end, we study the Larmor precession phenomenon. In both cases we show the theoretical and real experimental implementation. This article could be useful in introductory courses on quantum mechanics and nuclear magnetic resonance foundations, for example.
\end{abstract}

\maketitle 

\section{Introduction}

In traditional courses of quantum mechanics where we introduce the notion of single-spin dynamics, some interesting results are discussed in a theoretical way, without any experimental implementation or verification. On the other hand, simulating quantum physics is an interesting task that we can explore in the IBM Team's quantum plataform, namely, the IBM Quantum Experience (IBM-QE)~\cite{IBMQE}. From such platform we have access to a five-qubit system~\cite{Schumacher:95} in which we can manipulate it from unitary operations (unitary operators) and measurements. This platform has been used to experimentally confirm some theoretical results on quantum information and computation~\cite{Berta:16,Garcia:18,Rundle:17,Dueck:18,Das:18} as well as to implement high technology quantum protocols like teleportation~\cite{Santos:17,Fedortchenko:16}, among others~\cite{Hegade:19,Mahanti:19,Doronin:20,Devitt:16,Zhukov:18,Viyuela:18}.

Methods to circumvent experimental limitations of teaching quantum physics have been considered in literature through simulations~\cite{Merril:72,Benenti:08,Grivet:93}, experimental proposal in quantum optics~\cite{Skaar:04}, nuclear magnetic systems~\cite{Havel:02} and classical implementation of quantum computers~\cite{Vianna:18}. By using a different approach, in this paper we discuss how useful the IBM-QE can be in a physics teaching scenario, where we develop a didactic strategy to deal with the problem of experimentally present foundations of single-spin dynamics in presence of a magnetic field, including time-dependent fields. To this end, we first present the relevant elements that we need to know before starting the simulation of the examples we will present here. We discuss about the unitary operators that we will use in our simulations and how we can implement measurements in IBM-QE. By the way, we adopt an approach which allows us to introduce how measurement of some physical quantities are performed in a Nuclear Magnetic Resonance (NMR) experimental setup~\cite{Sarthour:Book}. Then, we discuss about the first interesting phenomena of single-spin dynamics, namely \textit{spin resonance} phenomena, associated with a spin-$\frac{1}{2}$ particle in presence of a strong static field along $z$-direction and a rotating magnetic field. We discuss how the resonance phenomenon emerges from a suitable choice of the rotating magnetic field frequency and how it allows us to promote transitions between ``spin up" and ``spin down" states of the system, even when such rotating field is weakly interacting with the system~\cite{Sarthour:Book,Nielsen:Book}. To conclude, we study the Larmor spin precession behavior.


\section{The IBM Quantum Experience}

\begin{figure}[t!]
	\centering
	\subfloat[]{\includegraphics[scale=0.58]{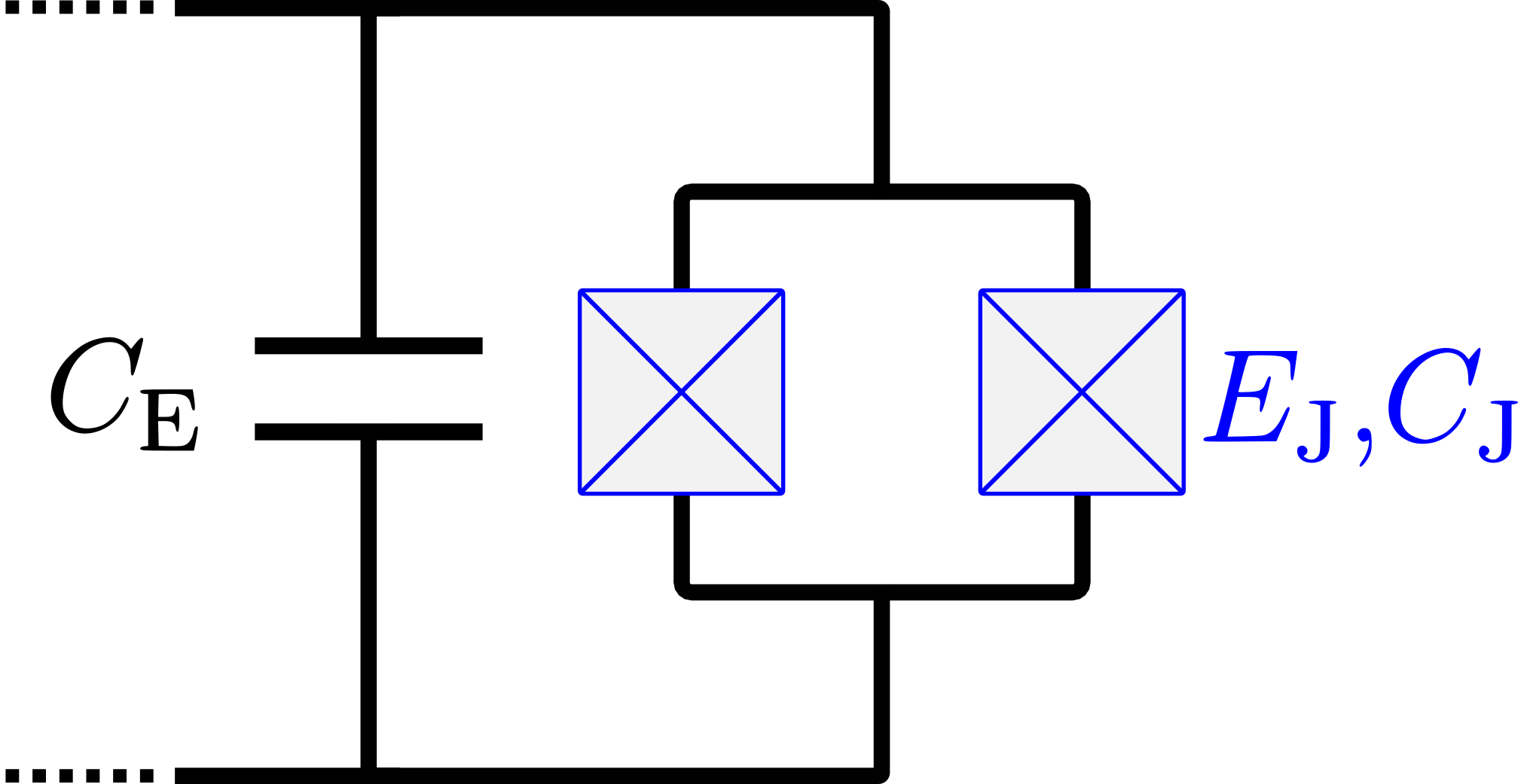} \label{FigCircSuper}} \quad 
	\subfloat[]{\includegraphics[scale=0.58]{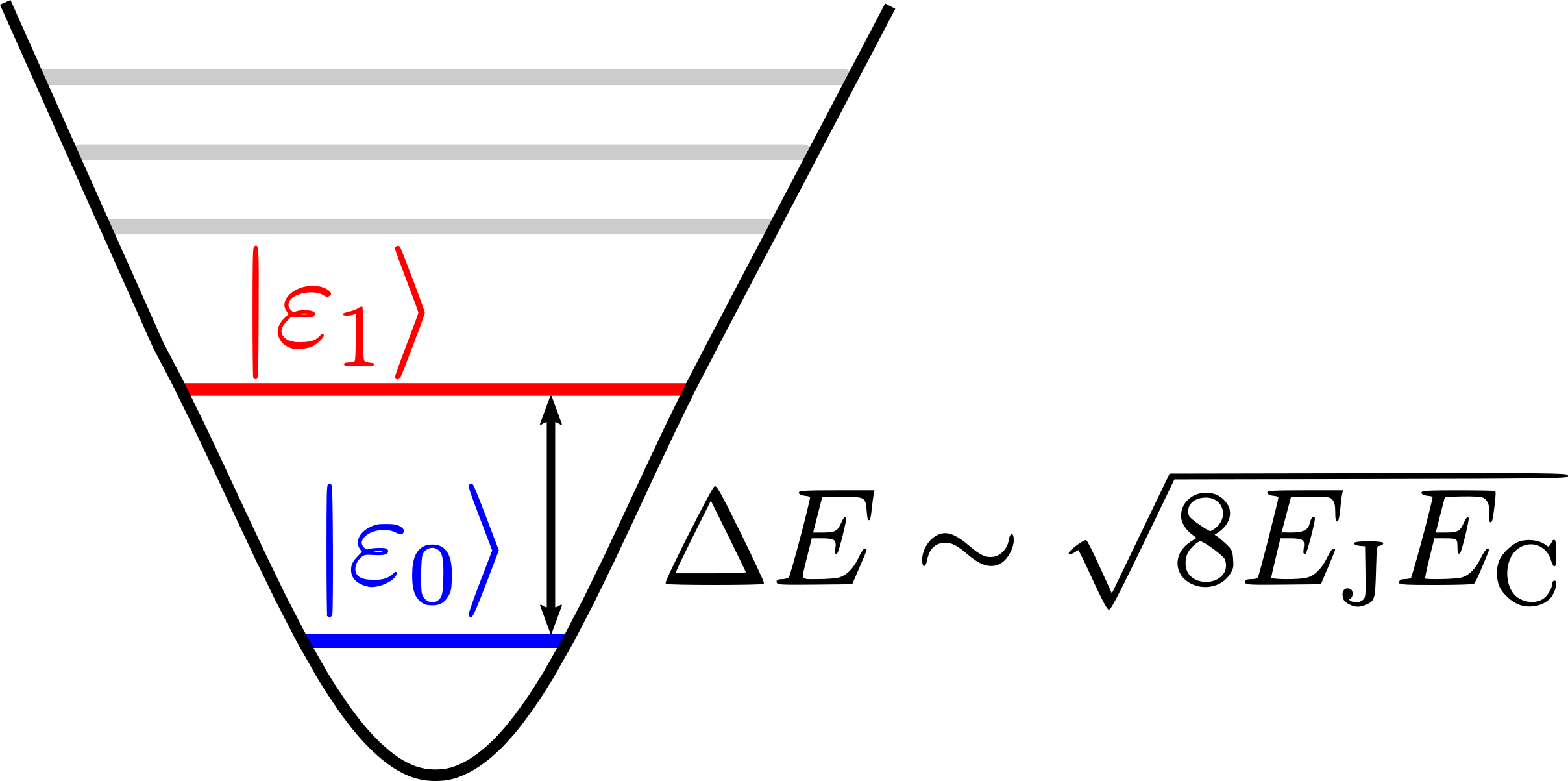} \label{FigEnergy}}
	\caption{\eqref{FigCircSuper} Sketch of the superconducting circuit that constitutes a transmon superconducting qubit, where we have two Josephson junctions with Josephson energy $E_{\text{J}}$ and capacitive energy $E_{\text{C}}$, connected to a external capacitance $C_{\text{E}}$. \eqref{FigEnergy} The representation (out of scale) of the energy levels of the system, where we highlight the energy levels used as the physical qubit.}
\end{figure}

The five-qubit IBM quantum chip is composed by five superconducting transmon qubits operating at a temperature scale around $15$~mK~\cite{FootSee1,Koch:07,Schreier08}. A superconducting transmon qubit is composed of two coupled Josephson junctions, with Josephson energy $E_{\text{J}}$ and capacitance $C_{\text{J}}$~\cite{Makhlin:01}, and shunted by an additional external capacitance $C_{\text{E}}$, as shown in Fig.~\ref{FigCircSuper}. As sketched in Fig.~\ref{FigEnergy}, this system presents a quantized energy levels structure $E_{n}$, where two of these energy levels (the two lowest-energy levels) are used as an artificial qubit for quantum computation, since quantum coherence is observed in such systems~\cite{Paik:11}. Because the Josephson junctions are superconducting devices, the system is composed by a large number of Cooper pairs (a pair of electrons), where the collective effect of these pairs allows us to use a superconducting qubit as ``artificial atom". In particular, we can use it as a two-level system. For this reason, this system is not a genuine two-level quantum system, like spin-$1/2$ particle in presence of a magnetic field or the polarization states of light. However, the characteristics and dynamics of superconducting qubits allows us to simulate such systems. For more relevant details about the superconducting transmon qubits, we recommend the Refs~\cite{Koch:07,Schreier08,Paik:11}.

In particular, we can perform simulations and real experimental realizations from two different quantum chips, known as the IBM Q 5 Tenerife (ibmqx4) and IBM Q 5 Yorktown (ibmqx2) quantum chips. Particularly, throughout this article we implement the experiments on IBM Q 5 Yorktown quantum chip, due to the good decoherence time scale of its qubits~\cite{IBMQE}. In this section we discuss on some important elements of the IBM-QE to be considered in this paper, but more information about IBM-QE quantum chips can be obtained from a number of papers in literature~\cite{Santos:17}, as well as from the own IBM-QE team beginners guide~\cite{IBMQE:Guide}.

\subsection{IBM-QE single-qubit operations}

Single qubit operations are unitary transformations on two level systems which, in a realistic physical scenario, arise as result of a dynamics provided by Schrödinger equation
\begin{eqnarray}
i\hbar \ket{\dot{\psi}(t)} = H(t) \ket{\psi(t)} \text{ , }
\end{eqnarray}
for some time dependent (or time independent) Hamiltonian $H(t)$. In particular, single-qubit operations can be obtained from different quantum evolutions, as in spin dynamics~\cite{Havel:02,Scarani:98} or in a quantum particle in a double-well~\cite{Foot:11}, for example. In this paper, the relevant single-qubit operations implementable by the IBM five-qubit chip are
\begin{align}
U_{1}(\lambda) &= \begin{bmatrix}
	1 & 0\\
	0 & e^{i\lambda}
\end{bmatrix} \text{, } 
U_{2}(\lambda,\phi) = \begin{bmatrix}
	\frac{1}{\sqrt{2}} & -\frac{e^{i\lambda}}{\sqrt{2}}\\
	\frac{e^{i\phi}}{\sqrt{2}} &  \frac{e^{i(\lambda+\phi)}}{\sqrt{2}}
\end{bmatrix} \text{, } 
U_{3}(\theta,\phi,\lambda) = \begin{bmatrix}
\cos\left( \frac{\theta}{2} \right) & -e^{i\lambda}\sin\left( \frac{\theta}{2} \right)\\
e^{i\phi}\sin\left( \frac{\theta}{2} \right) & e^{i(\lambda+\phi)}\cos\left( \frac{\theta}{2} \right)
\end{bmatrix} \text{, } \label{Us}
\end{align}
where $\lambda$, $\phi$ and $\theta$ are free real parameters that can be adjusted in accordance with the gate to be implemented. It is possible to see that the gates $U_{1}(\lambda)$ and $U_{2}(\lambda,\phi)$ can be obtained from gate $U_{3}(\theta,\phi,\lambda)$, but under experimental viewpoint the gates $U_{1}(\lambda)$ and $U_{2}(\lambda,\phi)$ can be more efficient than its counterpart obtained from $U_{3}(\theta,\phi,\lambda)$. In fact, each gate can be characterized from its run-time (time associated with real pulse sequence on transmon qubits), where the run-time for $U_{1}(\lambda)$ can be neglected, $U_{2}(\lambda,\phi)$ is a gate with duration of 1 unit of gate time and $U_{3}(\theta,\phi,\lambda)$ has run-time of 2 units of gate time. Therefore, it is most convenient to use the gates $U_{1}(\lambda)$ and $U_{2}(\lambda,\phi)$ whenever possible.

\subsection{Measuring physical quantities in a single-spin system}~\label{SecMeasurement}

The measurement process in physics depends on the system under consideration~\cite{Millhouse:13,Gerry:97,Scarani:98,Havel:02,Barman:19}. Here, we are interested in studying a single-spin-$\frac{1}{2}$ system driven by magnetic fields which act on it. Thus, since the system can be found in two different orthogonal states $\ketus$ and $\ketds$ associated with different values of the spin component, these states work as a computational basis in a Nuclear Magnetic Resonance based quantum computer~\cite{Scarani:98,Havel:02}. In particular, the IBM-QE allows us to implement measurements in computational basis $\ket{0} \equiv \ketus$ and $\ket{1} \equiv \ketds$, i.e., given a quantum state $\ket{\psi} = a\ketus+b\ketds$, the measurement outcome provides values for $|a|^2$ and $|b|^2$. Thus, the physical quantities which we can measure in IBM-QE are constrained to a set of measurements that can be obtained from parameters $|a|^2$ and $|b|^2$. For example, if we want to measure the expected value of the spin along $z$ direction, we have
\begin{align}
\Mcal_{z} = \bra{\psi}S_{z}\ket{\psi} = \frac{\hbar}{2}\left(|a|^2 - |b|^2\right) \text{ , } \label{Mz}
\end{align} 
where $S_{x}$, $S_{y}$ and $S_{z}$ are the spin operators defined as
\begin{align}
S_{x} = \frac{\hbar}{2} \begin{bmatrix}
0 & 1 \\
1 & 0
\end{bmatrix} \text{ , \, } S_{y} = \frac{\hbar}{2} \begin{bmatrix}
0 & -i \\
i & 0
\end{bmatrix} \text{ , and \, } S_{z} = \frac{\hbar}{2} \begin{bmatrix}
1 & 0 \\
0 & -1
\end{bmatrix} \text{ .}
\end{align}

Thus, it is possible to compute $\Mcal_{z}$. In addition, if we want to compute the expected value of spin along $x$ direction, we need to compute $\Mcal_{x} = \bra{\psi}S_{x}\ket{\psi}$. It can be done if we define the operator
\begin{align}
H = \frac{1}{\sqrt{2}}\begin{bmatrix}
1 & 1 \\
1 & -1 
\end{bmatrix} \text{ , }
\end{align}
where one uses the relation $S_{x} = HS_{z}H$ to write
\begin{align}
\Mcal_{x} = \bra{\psi}HS_{z}H\ket{\psi} = \bra{\psi_{x}}S_{z}\ket{\psi_{x}} \text{ , } \label{Mx}
\end{align} 
with $\ket{\psi_{x}}=H\ket{\psi}$. In quantum computation, the operator $H$ is a Hadamard gate and it can be implemented on IBM-QE. In conclusion, it means we can measure $\Mcal_{x}$ if we implement a Hadamard gate on the state $\ket{\psi}$ and then we measure the output state in computational basis. In a similar way, the measurement of the quantity $\Mcal_{y} = \bra{\psi}S_{y}\ket{\psi}$ can be done by defining the rotation operator
\begin{align}
R_{x}(\xi) = \begin{bmatrix}
\cos(\xi/2) & -i \sin(\xi/2) \\
-i \sin(\xi/2) & \cos(\xi/2)
\end{bmatrix} \text{ , }
\end{align}
that represents a rotation of an angle $\xi$ around $X$-direction~\cite{Nielsen:Book}. Thus, if we implement the operation with $\xi = -\pi/2$ we get
\begin{align}
R_{x}(-\pi/2) = \frac{1}{\sqrt{2}}\begin{bmatrix}
1 & i \\
i & 1 
\end{bmatrix} \text{ , }
\end{align}
so that we can show that $S_{y} = R_{x}(-\pi/2) S_{z} R_{x}^{\dagger}(-\pi/2)$, then
\begin{align}
\Mcal_{y} = \bra{\psi}S_{y}\ket{\psi} = \bra{\psi} R_{x}(-\pi/2) S_{z} R_{x}^{\dagger}(-\pi/2) \ket{\psi} = \bra{\psi_{y}} S_{z} \ket{\psi_{y}} \text{ , } \label{My}
\end{align} 
with $\ket{\psi_{y}} = R_{x}^{\dagger}(-\pi/2) \ket{\psi}$. Thus, we can measure any spin component along the directions $x$, $y$ and $z$. In addition, we can perform measurement of any physical quantity $\Omath$ which can be written as $\Omath = R_{\hat{r}}(\phi) S_{z} R^{\dagger}_{\hat{r}}(\phi)$, where $R_{\hat{r}}(\phi)$ denotes a rotation of an angle $\phi$ around direction $\hat{r}$, from equation
\begin{eqnarray}
\Ocal = \bra{\psi} \Omath \ket{\psi} = \bra{\psi} R_{\hat{r}}(\phi) S_{z} R^{\dagger}_{\hat{r}}(\phi) \ket{\psi} \text{ . }
\end{eqnarray}

\section{Single-spin dynamics on IBM Quantum Experience}

The spin is an internal degree of freedom of the electron that can be manipulated through external magnetic fields. In particular, for a spin-$\frac{1}{2}$ particle we have two distinct states $\ketus$ and $\ketds$. These two states satisfy the eigenvalue equation
\begin{eqnarray}
S_{z} \ketus = \frac{\hbar}{2} \ketus \quad \text{ and } \quad S_{z} \ketds = -\frac{\hbar}{2} \ketds \text{ , }
\end{eqnarray}
in which $\hbar$ is the reduced Planck's constant and $S_{n} = (\hbar/2) \sigma_{n}$, where $\sigma_{n}$ ($n=\{x,y,z\}$) denotes the Pauli matrices for a two-level system given by (with $i = \sqrt{-1}$)
\begin{eqnarray}
\sigma _{x} = \begin{bmatrix}
0 & 1\\
1 & 0
\end{bmatrix} \text{ , } \quad
\sigma _{y} = \begin{bmatrix}
0 & -i\\
i & 0
\end{bmatrix} \text{ and } \quad
\sigma _{z} = \begin{bmatrix}
1 & 0\\
0 & -1
\end{bmatrix} \text{ , }
\end{eqnarray}
so that the matrix form for the basis $\ketus$ and $\ketds$ reads as $\ketus = [\text{ } 1 \quad 0\text{ } ]^{\text{t}}$ and $\ketds = [\text{ } 0 \quad 1\text{ } ]^{\text{t}}$, with superscript ``t" denoting the transpose of a matrix.

\subsection{Simulating Nuclear Magnetic Resonance phenomenon}

Let us consider the coupling dynamics of a single spin-$\frac{1}{2}$ with a rotating magnetic field $\vec{B}(t)$ given by
\begin{align}
\vec{B}(t) = \vec{B}_{0} + \vec{B}_{\text{rf}}(t) \text{ , }
\end{align}
where $\vec{B}_{0} = B_{0}\hat{z}$ is the field component along the $Z$-direction and $\vec{B}_{\text{rf}}(t)$ is a time-dependent transverse magnetic fields, given by
\begin{align}
\vec{B}_{\text{rf}}(t) = B_{\text{rf}} \left[ \cos (\omega t) \hat{x} + \sin (\omega t) \hat{y}\right]  \text{ , }
\end{align}
where $\omega$ is the rotating frequency of the transverse magnetic field, also known as Radio-Frequency (rf) field. Therefore, the Hamiltonian of the system is written as
\begin{equation}
H\left( t\right) = \frac{\hbar\omega _{0}}{2%
}\sigma _{z}+\frac{\hbar\omega _{\text{rf}}}{2}\left[ \cos \left( \omega t\right) \sigma
_{x}+\sin \left( \omega t\right) \sigma _{y}\right] \text{ ,}
\label{Hamiltonian}
\end{equation}
where $\omega _{0}$ is the Larmor frequency and $\omega _{\text{rf}}$ is the coupling intensity between the spin and the rf-field. The solution of the Schrödinger equation for the system is known and given by~(See appendix~\ref{ApSpinSol})
\begin{equation}
\ket{\psi \left( t\right)} =\exp \left[ -i\frac{\omega }{2} t\sigma _{z}\right] \exp \left[ -\frac{i}{\hbar }\tilde{H}t\right] \ket{\psi \left( 0\right)} \text{ , } \label{SingSpinDyn}
\end{equation}
where
\begin{align}
\tilde{H} = \hbar \frac{\omega _{0}-\omega }{2}\sigma
_{z}+ \hbar\frac{\omega _{\text{rf}}}{2}\sigma _{x} \text{ . }
\end{align}

In matrix form, we write each evolution operator as
\begin{align}
U_{0}(t) &= \exp \left[ -i\frac{\omega }{2}t\sigma _{z}\right] = e^{-\frac{i\omega t}{2}} \begin{bmatrix}
1 & 0\\
0 & e^{i\omega t}
\end{bmatrix} \text{ , } \label{U0} \\
U_{xz}(t) &= \exp \left[ -\frac{i}{\hbar }\tilde{H}t\right] = \begin{bmatrix}
\cos \left(\frac{\Omega t}{2}\right) + i\frac{\Delta}{\Omega}\sin \left(\frac{\Omega t}{2}\right) & i\frac{\omega_{\text{rf}}}{\Omega}\sin \left(\frac{\Omega t}{2}\right)\\
i\frac{\omega_{\text{rf}}}{\Omega}\sin \left(\frac{\Omega t}{2}\right) & \cos \left(\frac{\Omega t}{2}\right) - i\frac{\Delta}{\Omega}\sin \left(\frac{\Omega t}{2}\right)
\end{bmatrix} \text{ , }
\end{align}
where $\Delta = \omega_{0} - \omega$ is a detuning between the rf-field and the Larmor frequencies, and $\Omega^2 = \Delta^2+\omega_{\text{rf}}^2$ is the effective Rabi frequency, which takes into account the effects due to the detuning~\cite{Hu:18}. It is worth to mention that the term $e^{-\frac{i\omega t}{2}}$ in $U_{0}(t)$ represents a global phase and can be neglected, as we will do from now on. In particular, here we will simulate the resonance phenomenon, where a small rf-field ($|\omega_{\text{rf}}| \ll |\omega_{0}|$) can be used to flip the spin state when the rf-field oscillation frequency is close to $\omega_{0}$ ($|\Delta| \rightarrow 0$). Under this configuration, the operator $U_{xz}(t)$ becomes
\begin{align}
U_{xz}^{\text{ress}}(t) &= \begin{bmatrix}
\cos \left(\frac{\omega_{\text{rf}} t}{2}\right) & i\sin \left(\frac{\omega_{\text{rf}} t}{2}\right)\\
i\sin \left(\frac{\omega_{\text{rf}} t}{2}\right) & \cos \left(\frac{\omega_{\text{rf}} t}{2}\right) 
\end{bmatrix} \text{ , }
\end{align}
and the system will evolve as
\begin{equation}
\ket{\psi ^{\text{ress}}(t)} = U_{0}(t)U_{xz}^{\text{ress}}(t) \ket{\psi \left( 0\right)} \text{ . } \label{SingSpinDynRess}
\end{equation}

In case where the initial state is given by $ \ket{\psi \left( 0\right)} = \ketus$, we compute the time-dependence of the $Z$-spin component and we find
\begin{eqnarray}
\Mcal_{z}(t) = \bra{\psi(t)} S_{z} \ket{\psi(t)} = \frac{\hbar}{2} \cos (\omega_{\text{rf}} t) \text{ . }
\end{eqnarray}

Notice that the parameter $\omega$ does not develop any role in the above quantity. We choose this particular state because the standard input state in IBM quantum chip is the computational state $\ket{0}$, namely the spin-up state $\ketus$.

To simulate the spin dynamics of the Eq.~\eqref{SingSpinDynRess} we need to map the evolution operators $U_{0}(t)$ and $U_{xz}^{\text{ress}}(t)$ into the available parameters in IBM-QE. We mean, from the Eqs.~\eqref{Us} we need to identify the operator $U_{0}(t)$ as $U_{1}(\lambda_{1})$ and the operator $\Ucal_{\text{ress}} (t)$ as $U_{3}(\theta,\phi,\lambda)$ for a particular choice of the parameters $\lambda$, $\phi$ and $\theta$. A first point to be highlighted is that $\lambda$, $\phi$ and $\theta$ are dimensionless parameters, so we need to get a correspondence between the dimensionless parameters of $U_{0}(t)$ and $U_{xz}^{\text{ress}}(t)$ with the parameters associated to the operators $U_{1}(\lambda_{1})$ and $U_{3}(\theta,\phi,\lambda)$, respectively. In our case, it is possible to see that $U_{1}(\omega t) = U_{0}(t)$ and $U_{3}(\frac{\omega_{\text{rf}} t}{2},\frac{\pi}{2},\frac{3\pi}{2}) = \Ucal_{\text{ress}} (t)$. In conclusion, $\Ucal_{\text{ress}} (t)$ would be simulated through the sequence
\begin{eqnarray}
\Ucal^{\text{sim}}_{\text{ress}} (t) = U_{1}(\omega t)U_{3}\left(\frac{\omega_{\text{rf}} t}{2},\frac{\pi}{2},\frac{3\pi}{2}\right) \text{ , }
\end{eqnarray}

\begin{figure*}[t!]
	\centering
	\subfloat[]{\includegraphics[scale=0.22]{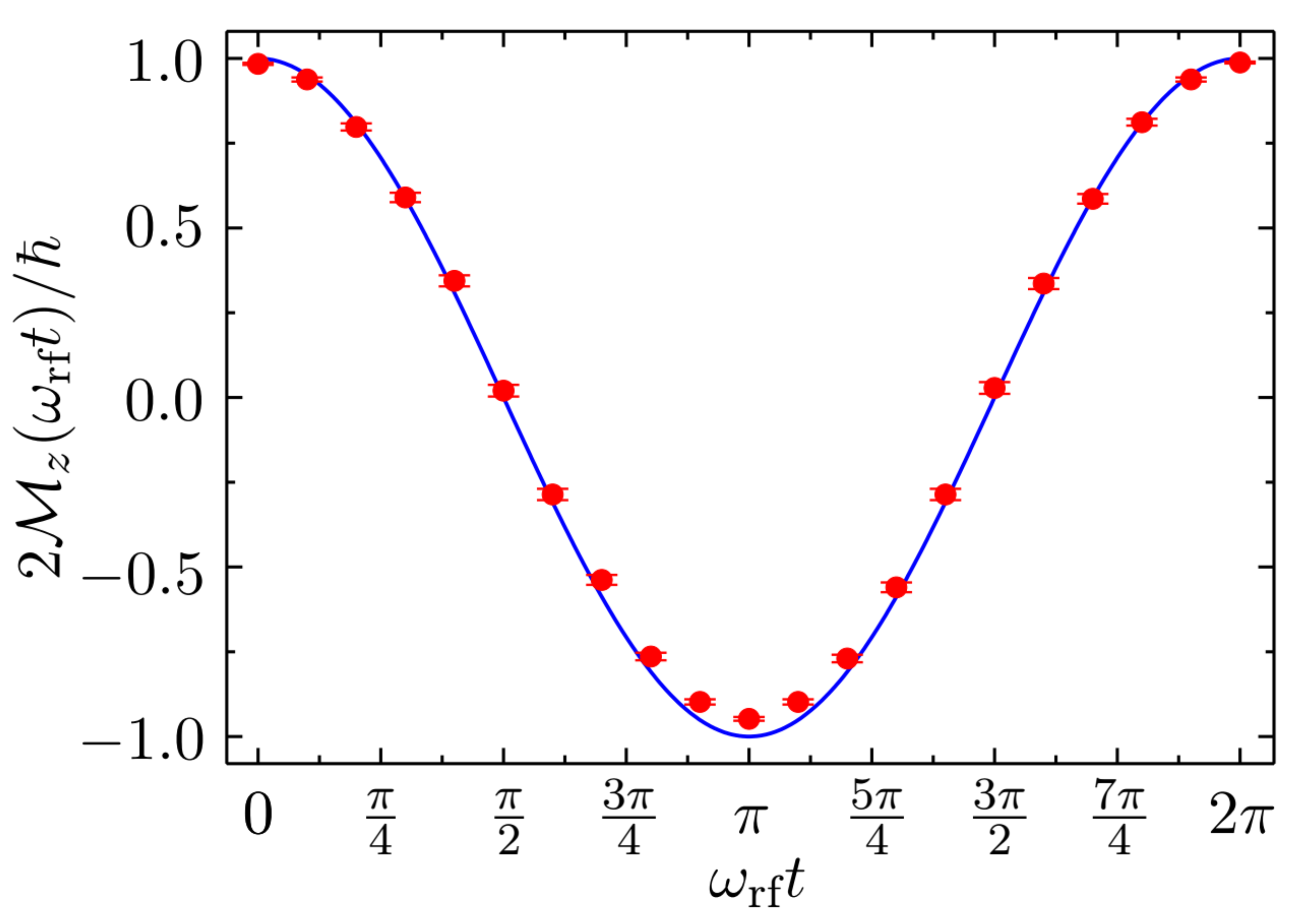}\label{GraphSpinDyn}} \quad
	\subfloat[]{\includegraphics[scale=0.22]{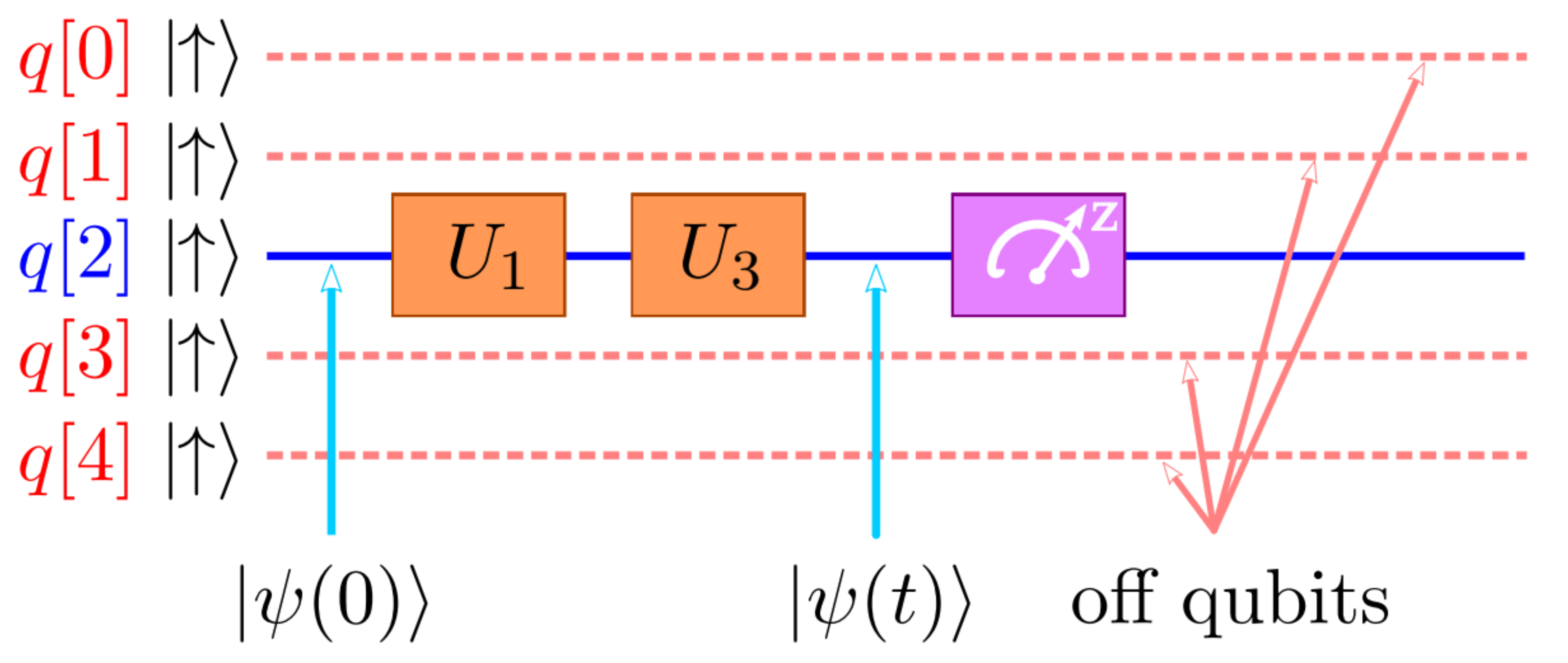}\label{CircuitResonance}}
	\caption{\eqref{GraphSpinDyn} Theoretical (line) and experimental data (dots) of the single spin measurement at resonance (as multiple of $\hbar / 2$). In our experiment we set $N = N_{\text{max}} = 8192$~shots. \eqref{CircuitResonance} Circuit implemented on IBM Q 5 Yorktown chip (ibmqx2), where we highlight the qubit used in experiment (continuum blue line) while the rest of the qubit were disregarded during the experiment (dashed light red lines). The magenta box represents a measurement in computational basis.}
	\label{FigResonance}
\end{figure*}

\begin{table}[t!]
	\caption{Physical parameters obtained from last calibration before the experimental implementation shown in Fig.\ref{FigResonance}. For this experiment, the calibration date is 2019-05-24 08:04:12~AM. See Appendix~B for more details about the physical meaning of each quantity in this table.}
	\begin{tabular}{r|cccccc}
		Parameters                & & $q[0]$ & $q[1]$ & $q[2]$ & $q{[3]}$ & $q[4]$ \\ \hline
		Frequency (GHz)           & & 5.29   & 5.24   & 5.03   & 5.3        & 5.08   \\
		T1 ($\mu$s)               & & 53.40  & 62.10  & 65.50  & 60.90      & 49.50  \\
		T2 ($\mu$s)               & & 42.20  & 54.40  & 57.20  & 28.90      & 59.90  \\
		Gate error ($10^{-3})$    & & 3.35   & 1.55   & 4.64   & 3.44       & 5.84   \\
		Readout error ($10^{-2})$ & & 4.80   & 24.20  & 1.70   & 1.70       & 32.00 
	\end{tabular}
	\label{TableResonance}
\end{table}

The quantum circuit that executes this task is shown in Fig.~\ref{FigResonance}. In Fig.~\ref{GraphSpinDyn} we present the theoretical and experimental data of the spin dynamics for the case where $\omega = \omega_{0}$ (resonance situation). To this end, we encode our spin (to be driven by Hamiltonian in Eq.~\eqref{Hamiltonian}) as the qubit $q[2]$ shown in Fig.~\ref{CircuitResonance}. The experimental chip condition is shown in table~\ref{TableResonance}, as provided by IBM team before we start the experimental procedure. In table~\ref{TableResonance}, we present the frequency $\nu$ of the energy gap between states $\ketds$ and $\ketus$ of the two-level system used in experiment, namely, $\nu = (\Delta E/\hbar)$, where $\Delta E$ is the difference of energy between excited state and fundamental one. 
In addition, it is shown the relevant quantities $T_{1}$ and $T_{2}$ associated with the natural undesired (decoherence) effects acting on the system (for more details about them, see Sec.~\ref{DecoherenceAp} fo the Appendix~\ref{NonUnitaryAp}). Moreover, we present the gate and readout errors in the table. These undesired effects are different from the decoherence ones, in particular the gate error arises from the imprecise control of the pulses used to manipulate the qubit, while readout error takes into account systematic errors and experimental limitations of the evaluation process to measure the spin components of the qubit. A brief discussion on gate and readout errors is considered in Sec.~\ref{ErrorAp} of the Appendix~\ref{NonUnitaryAp}.

After implementing the unitary operations (which simulate the dynamics), we measure the $Z$-spin component in order to see some spin flip in the system (as expected in resonance situation). Up to a small experimental error, the experimental results show a good compatibility with the theoretical expected values. The error bar is computed from relation~\cite{FootSee2} $\Delta p = \pi \sqrt{p_{0}(1-p_{0})/N}$, where $N$ is the number of experimental shots and $p_{0} = |\interpro{\uparrow\!}{\psi}|^2$. It is important to justify here the choice of the qubit $q[2]$ as our working system. As discussed in Appendix~\ref{NonUnitaryAp}, the scale $T_{1}$ and $T_{2}$ times shown in Table~\ref{TableResonance} are related with the timelife of the coherence in system, so the quantum gates should be implemented during a computation run time $T_{\text{comp}} \ll T_{1}$ and $T_{\text{comp}} \ll T_{2}$. Typically, the time required to implement each gate is of the order of $10$~ns, so the circuits used in this manuscript are implemented within a computation run time of the order of few tens of nanoseconds~\cite{Wendin:17,Barends:14}. Consequently, in our particular experiment we do not need to worry with the decohering effects acting on the system. On the other hand, the gate and readout errors are important for our discussion. In order to minimize these errors in our experiment, we used the physical qubit $q[2]$ because it has the best (smallest) gate and readout errors. Eventually, one obtains a different result if one chooses another option. In particular, if we choose the physical qubit $q[4]$, it is not expected a good compatibility of our results with the theoretical expected values.

\subsection{Spin precession}

Now, as a second application, we discuss the experimental simulation of the Larmor precession phenomenon. Let us consider the system at the initial state
\begin{eqnarray}
\ket{\psi_{0}} = \cos \left(\frac{\theta}{2}\right) \ketus + \sin \left(\frac{\theta}{2}\right) \ketds \text{ , }
\end{eqnarray}
where $\theta \in [0,\pi]$. If the system is driven by a time-independent magnetic field along $z$-direction, i.e., $\vec{B}_{0} = B_{0}\hat{z}$, the associated Hamiltonian is read as
\begin{align}
H_{0} = \frac{\hbar \omega_{0}}{2} \sigma_{z} \text{ , }
\end{align}
the evolved state of the system will be given by (up to a global phase)
\begin{align}
\ket{\psi (t)} = \tilde{U}_{0}(t) \ket{\psi_{0}} = \cos \left(\frac{\theta}{2}\right) \ketus + e^{i\omega t} \sin \left(\frac{\theta}{2}\right) \ketds \text{ , }
\end{align}
where $\tilde{U}_{0}(t) = e^{-\frac{i}{\hbar} H_{0} t} = \text{diag} [ \text{ } 1 \quad e^{i\omega_{0} t} \text{ } ]$ is the evolution operator. Now, by computing the behavior of the physical quantities defined in Eqs.~\eqref{Mz},~\eqref{Mx} and~\eqref{My}, we get
\begin{align}
\Mcal_{z} = \frac{\hbar}{2}\cos (\theta) \text{ , } \quad \Mcal_{x} (t) = \frac{\hbar}{2}\sin(\theta) \cos (\omega_{0} t) \text{ , } \quad \Mcal_{y} (t) = \frac{\hbar}{2}\sin(\theta)\sin (\omega_{0} t) \text{ . } \label{Magnetization}
\end{align}
 
In order to see the precession behavior, if we define the spin vector $\vec{\Mcal}_{xy}$ in $xy$-plane as
\begin{eqnarray}
\vec{\Mcal}_{xy} = \Mcal_{x}\hat{x} + \Mcal_{y}\hat{y} \text{ , }
\end{eqnarray}
so that we use the Eqs.~\eqref{Magnetization} and we get
\begin{eqnarray}
\vec{\Mcal}_{xy} =  \frac{\hbar}{2} \sin(\theta) \left[ \cos (\omega_{0} t)\hat{x} + \sin (\omega_{0} t)\hat{y} \right] \text{ , }
\end{eqnarray}

\begin{figure*}
	\centering
	\subfloat[]{\includegraphics[scale=0.22]{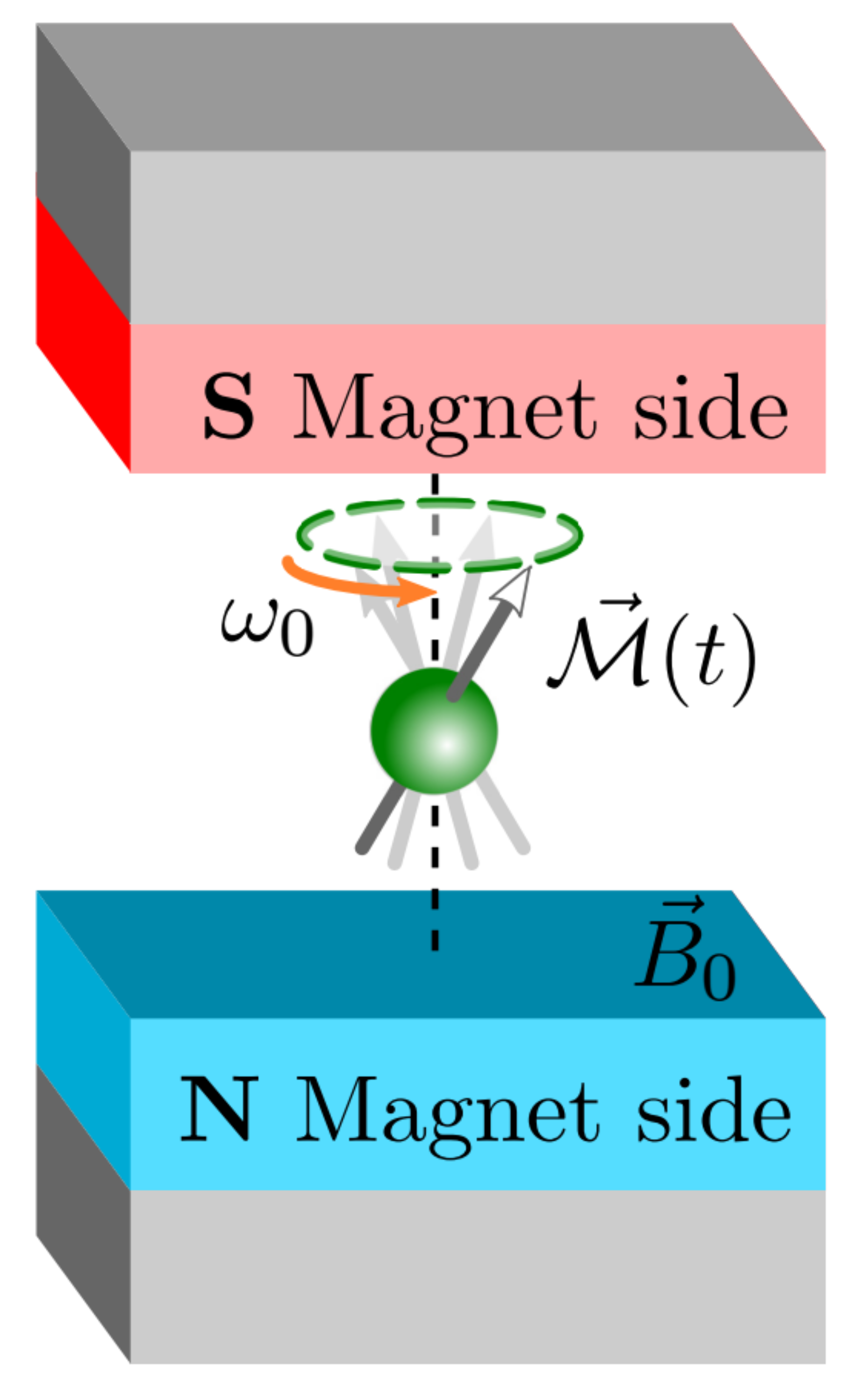}\label{Magnet}} \quad\quad
	\subfloat[]{\includegraphics[scale=0.22]{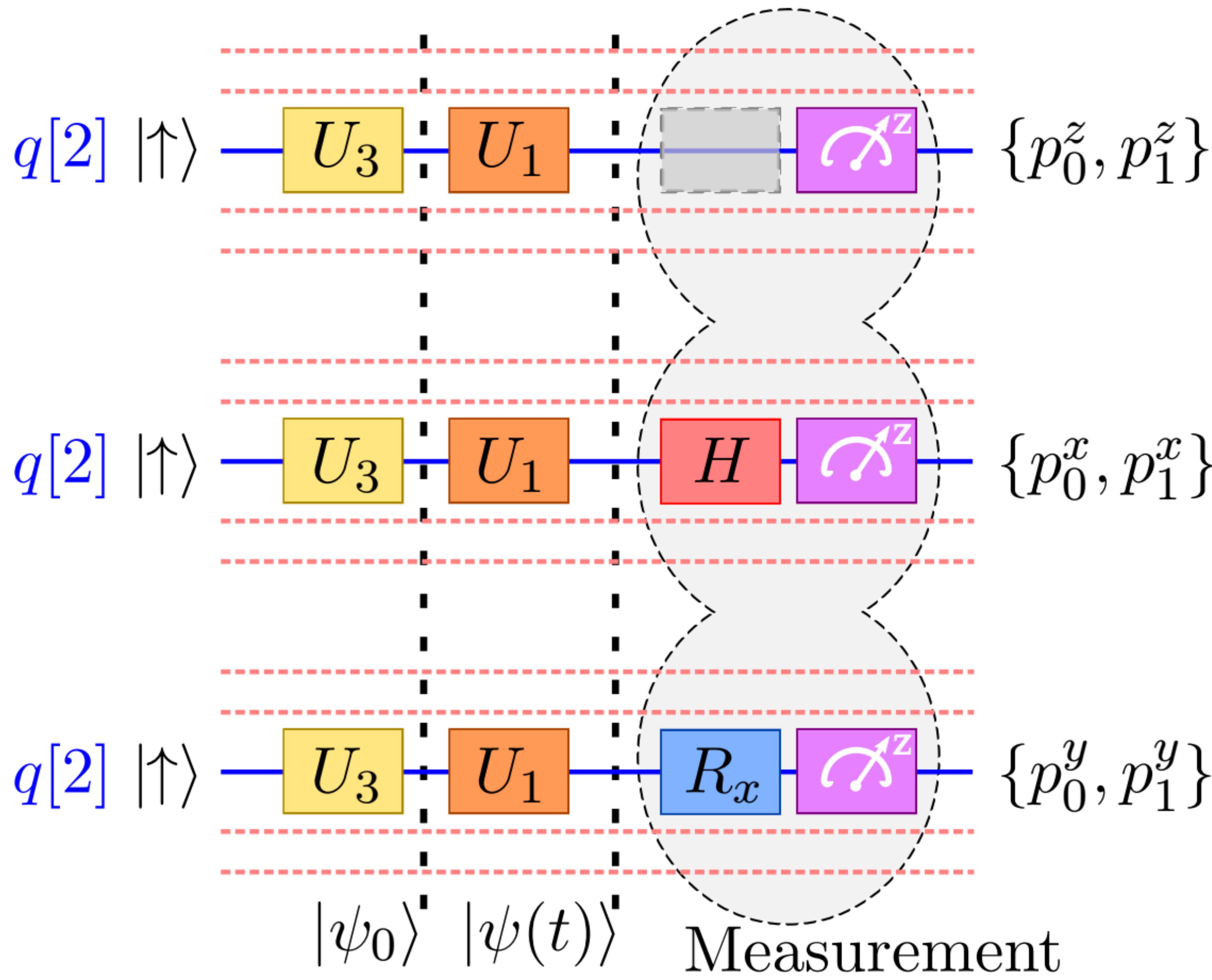}\label{CircuitPrecession}}
	\caption{\eqref{Magnet} Sketch of a single-spin inside a static magnetic field along $Z$-direction, where the precession behavior is highlighted. \eqref{CircuitPrecession} Circuit used for each dynamics and measurement implemented on IBM Q 5 Yorktown chip (ibmqx2), where we highlight the qubit used in experiment (continuum blue line) while the rest of the qubits are disregarded during the experiment (dashed light red lines). The magenta box represents a measurement in computational basis and provides us the probabilities $\{p^{x,y,z}_{0},p^{x,y,z}_{1}\}$.}
	\label{FigPrecession}
\end{figure*}

The geometrical representation of $\vec{\Mcal}_{xy}$ is a spinning vector around $z$-axis (in the $xy$-plane) and its norm is $||\vec{\Mcal}_{xy}|| = \hbar|\sin(\theta)|/2$. Therefore, when we define the spin vector $\vec{\Mcal} = \Mcal_{z}\hat{z} + \vec{\Mcal}_{xy}$ we get
\begin{align}
\vec{\Mcal}(t) =  \frac{\hbar}{2}\cos(\theta)\hat{z} +  \frac{\hbar}{2}\sin(\theta) \left[ \cos (\omega_{0} t)\hat{x} + \sin (\omega_{0} t)\hat{y} \right] \text{ , }
\end{align}
since $\Mcal_{z} = \hbar\cos(\theta)/2$. The geometrical representation of $\vec{\Mcal}(t)$ is shown in Fig.~\ref{Magnet}. For this reason, we call such dynamics \textit{Larmor spin precession} or \textit{spin precession}.

To simulate such dynamics, we need to prepare the initial input state $\ket{\psi_{0}}$, then we implement the evolution as provided by the evolution operator $\tilde{U}_{0}(t)$. Then, we measure the spin component along directions $x$, $y$ and $z$. Thus, the circuit which simulates the spin precession is shown in Fig.~\ref{FigPrecession} and we can build it following three steps:
\begin{itemize}
	\item[(1)] The initial state: As mentioned, the IBM-QE has a natural input state given by $\ket{0} = \ketus$, so any algorithm input state should be achieved from it. In particular, the initial state $\ket{\psi_{0}}$ can be obtained from $\ketus$ through the unitary operation
	\begin{align}
	U_{\text{inp}} = \begin{bmatrix}
	\cos\left( \frac{\theta}{2} \right) & -\sin\left( \frac{\theta}{2} \right)\\
	\sin\left( \frac{\theta}{2} \right) & \cos\left( \frac{\theta}{2} \right)
	\end{bmatrix} \text{. }
	\end{align}
	The above unitary operator is obtained from the gate $U_{3}(\theta,\phi,\lambda)$ in Eq.~\eqref{Us} if we set $\lambda = 0$ $\phi = 0$. Thus, we have $U_{\text{inp}} = U_{3}(\theta,0,0)$.
	\item[(2)] The evolution: As previously discussed, the evolution as provided by operator $\tilde{U}_{0}(t)$ can be achieved through the $U_{1}(\lambda)$ gate (up to a global phase) whenever we set $\lambda = \omega_{0} t$.
	\item[(3)] The measurement: The last circuit step is the measurement. To measure the $z$-spin component we do not need to implement rotations, since the natural measurement basis of the IBM-EQ is the computational basis (``spin-up " and ``spin-down" basis in our case). The spin measurement along $x$ and $y$ direction are obtained as discussed in Sec.~\ref{SecMeasurement}, so we need to implement a Hadamard ($H$) gate and $R_{x} = R_{x}^{\dagger}(-\pi/2)$ gate to measure in $x$ and $y$ basis, respectively.
\end{itemize}

\begin{table}[t!]
	\caption{Physical parameters obtained from last calibration before the experimental implementation shown in Fig.\ref{FigPrecessionGraph}. For this experiment, the last calibration date is 2019-05-29 08:07:25~AM. See Appendix~B for more details about the physical meaning of each quantity in this table.}
	\begin{tabular}{r|cccccc}
		Parameters                & & $q[0]$ & $q[1]$ & $q[2]$ & $q{[3]}$ & $q[4]$ \\ \hline
		Frequency (GHz)           & & 5.29   & 5.24   & 5.03   & 5.3        & 5.08   \\
		T1 ($\mu$s)               & & 21.90  & 51.10  & 70.10  & 60.30      & 50.40  \\
		T2 ($\mu$s)               & & 26.50  & 42.10  & 62.20  & 27.10      & 51.20  \\
		Gate error ($10^{-3})$    & & 9.28   & 1.46   & 4.21   & 3.95       & 3.44   \\
		Readout error ($10^{-2})$ & & 9.50   & 26.10  & 1.60   & 3.70       & 37.90 
	\end{tabular}
	\label{TablePrecession}
\end{table}

\begin{figure*}[t!]
	\centering
	\subfloat[]{\includegraphics[scale=0.22]{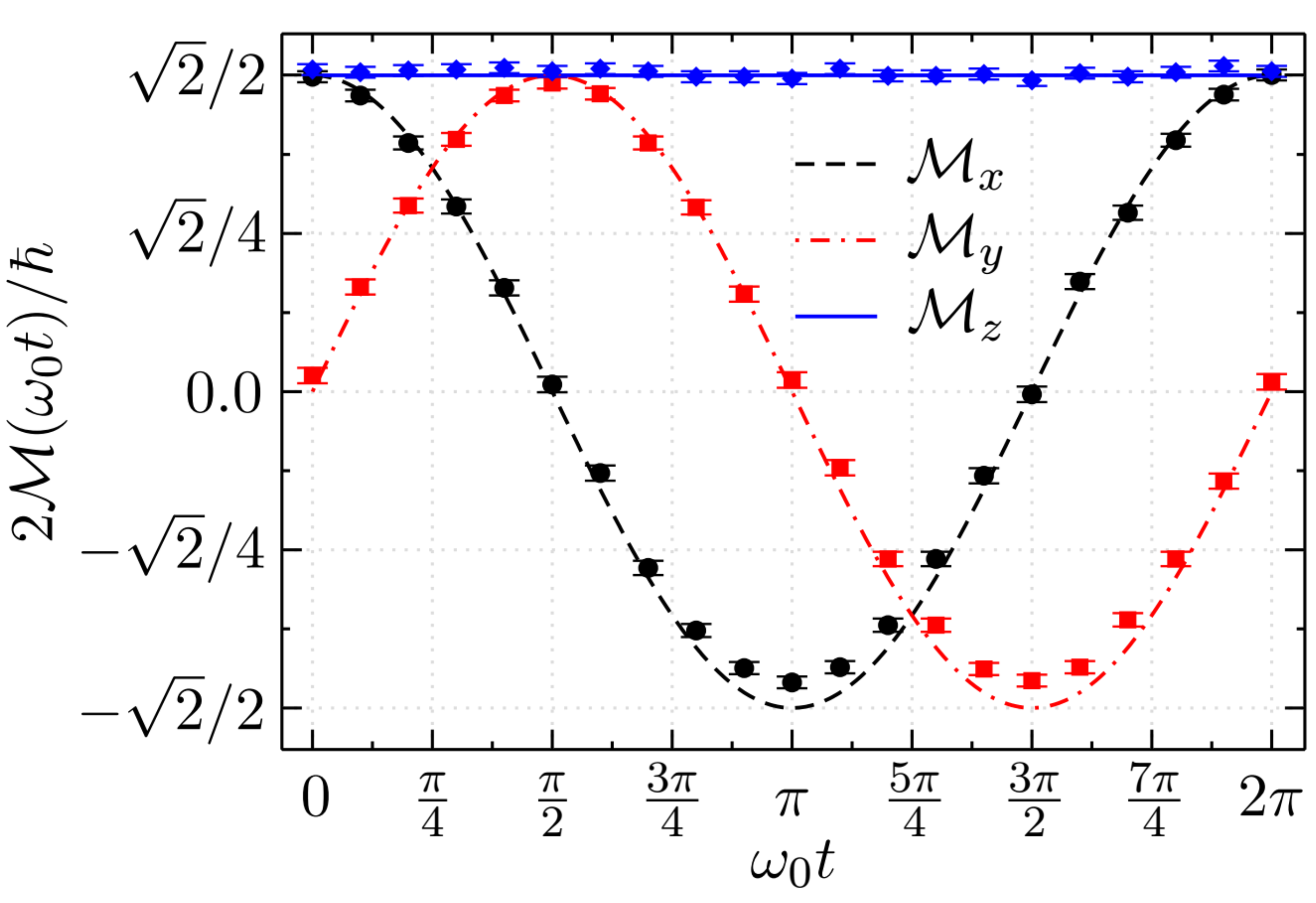}\label{Precession-pi-4}} \quad
	\subfloat[]{\includegraphics[scale=0.22]{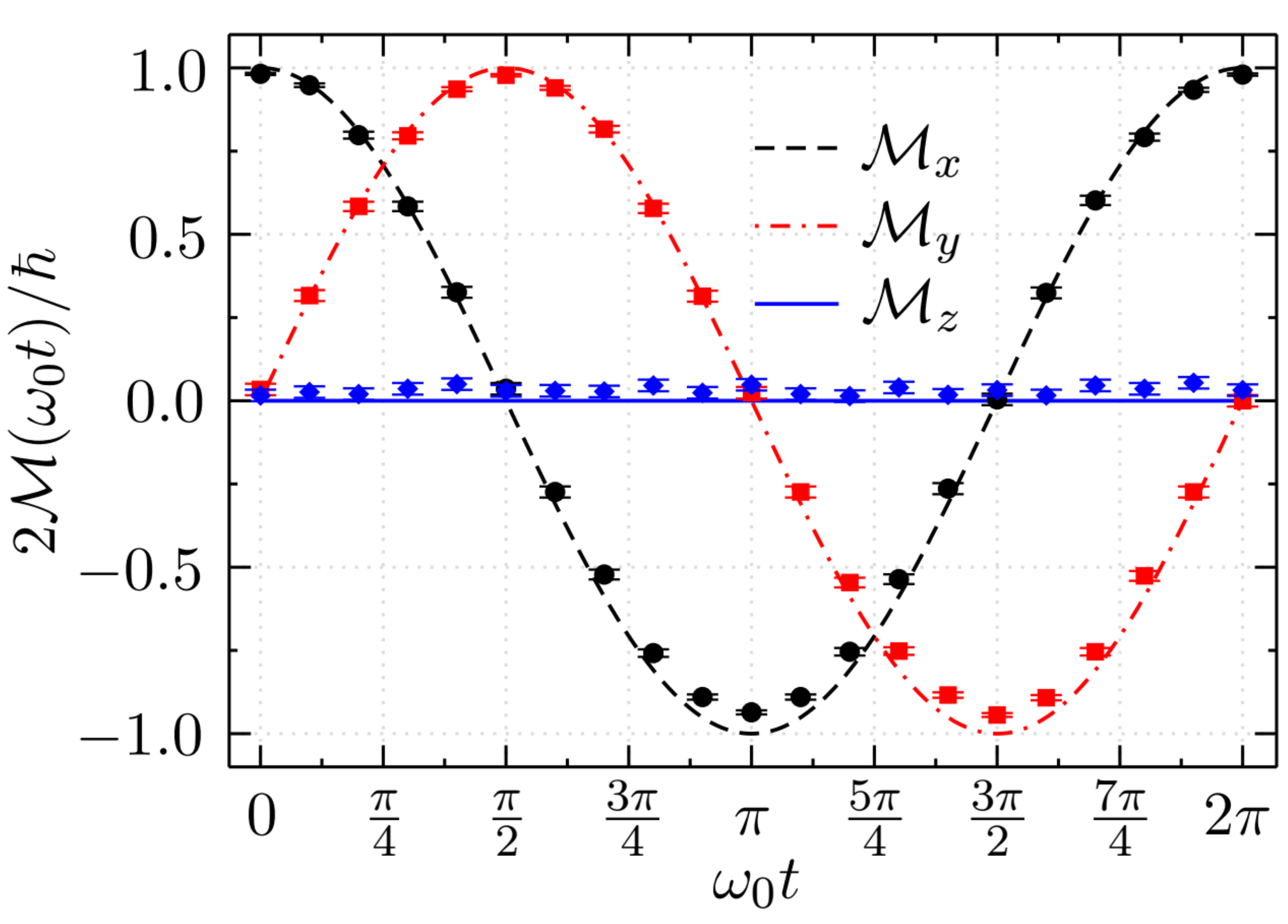}\label{Precession-pi-2}}
	\caption{Spin measurement $\Mcal$ (in multiple of $\hbar / 2$) along the direction $x$ (dashed black curves), $y$ (dashed  red dot curves) and $z$ (continuum blue curves) as function of $\omega_{0} t$ for the choices \eqref{Precession-pi-4} $\theta = \pi/4$ and \eqref{Precession-pi-2} $\theta = \pi/2$. In our experiment we set $N = N_{\text{max}} = 8192$~shots and we implement the circuit on IBM Q 5 Yorktown chip (ibmqx2). Points denote the experimental data obtained from IBM-QE platform.}
	\label{FigPrecessionGraph}
\end{figure*}

The experiment of the precession spin was implemented with same qubit as in previous experiment, but with new physical parameters as shown in Table~\ref{TablePrecession}. In experiment we have two dimensionless parameters to be adjusted, namely the parameter $\theta$ associated with initial state and the quantity $\omega_{0} t$ due to the time-dependence of the operator $\tilde{U}_{0}(t)$. Thus, we choose two values of $\theta$ and we experimentally study the behavior of the physical quantities defined in Eq.~\eqref{Magnetization} as function of $\omega_{0} t$. Since we need to start the protocol with the same state for each measurement, we set the same parameter $\theta$ in gate $U_{3}$ for each circuit of the Fig.~\ref{CircuitPrecession}, while we vary the parameter $\lambda$ of the gate $U_{1}$ used to encode the dimensionless value of $\omega_{0} t$. Given a same initial for all experiments, each circuits in figures provides us the probabilities values $p_{0}^{x,y,z}$ and $p_{1}^{x,y,z}$, then we obtain a set of experimental values of the quantities $\Mcal_{x,y,z}^{\text{exp}}(\omega_{0} t)$ from equation
\begin{eqnarray}
\Mcal_{x,y,z}^{\text{exp}}(\omega_{0} t) = \frac{\hbar}{2} \left[p_{0}^{x,y,z}(\omega_{0} t) - p_{1}^{x,y,z}(\omega_{0} t) \right] \text{ . }
\end{eqnarray}

Therefore, it allows us to compare the experimental result with theoretical values given in Eq.~\eqref{Magnetization} and the results are shown in Fig.~\ref{FigPrecessionGraph}. Again, we compute the error bar for each circuit from relation~\cite{FootSee2} $\Delta p^{x,y,z} = \pi \sqrt{p^{x,y,z}_{0}(1-p^{x,y,z}_{0})/N}$. Thus, we can see the good agreement between theoretical predictions and experimental data.

\section{Conclusion}

Here we have presented a didactic proposal which can be used to present a more consistent study on quantum single-spin dynamics, where we discussed how to remotely implement its experimental realization in a superconducting quantum processor. To this end, we encourage the use of the IBM-QE platform as a resource for teaching quantum mechanics from an experimental approach, in completeness with the theoretical one. Here we have shown the required elements (quantum gates and measurement) to study single-spin dynamics from IBM-QE systems. As a demonstration of how useful the IBM-QE platform can be, we discuss about two particular and interesting dynamics of single-spin system. As a first example, we have discussed about the resonance phenomena in systems composed of nuclear spins, with high applicability in nuclear magnetic resonance based quantum technologies. To conclude, we have studied the Larmor precession phenomena. Both applications could be experimentally verified. It is worth to highlight that both experiments were implemented in a relatively short time, for our case, in average, it took between 3 and 10 ~minutes for getting the experimental data file for each execution (each point of the graph).

The single-qubit dynamics studied in this paper can be implementable through different physical systems. For example, by using the photon polarization, a number of half-wave plate and quarter-wave plate can be used to simulate the dynamics. In this system, the readout process can be done through state tomography of the photon polarization. However, because this system needs a laboratory with high technology electronic devices, such laboratories are not accessible for a large number of universities. A second possibility of physical system, which can implement the experiment discussed here can be, is a NMR experimental setup. However, these system are orders of magnitude more expensive than optical ones. Therefore, the IBM-QE project is a possibility of experimental realizations on quantum information and computation for a number of universities around the world.

\begin{acknowledgments}
ACS acknowledges financial support from the Brazilian agencies Conselho Nacional de Desenvolvimento Científico e Tecnológico (CNPq), Brazilian National Institute of Science and Technology for Quantum Information [CNPq INCT-IQ (465469/2014-0)] and the Coordena\c{c}\~ao de Aperfei\c{c}oamento de Pessoal de N\'{\i}vel Superior (CAPES) (Finance Code 001). We are extremely grateful to IBM-QE project and to IBM-QE team for providing the experimental data used in this manuscript. To end, it is important to mention that the results and discussions expressed in this manuscript are the viewpoint of the authors and do not reflect the official policy or position of IBM or the IBM-QE team. 
\end{acknowledgments}

\appendix

\section{Exact solution for spin-$\frac{1}{2}$ dynamics in a rotating magnetic field}\label{ApSpinSol}

In this Appendix we discuss about the exact solution of the spin dynamics of a particle in presence of a time-dependent magnetic field given by $\vec{B}(t) = \vec{B}_{0} + \vec{B}_{\text{rf}}(t)$, where $\vec{B}_{0} = B_{0}\hat{z}$ and $\vec{B}_{\text{rf}}(t)$ are a static magnetic field along $Z$-direction and a time-dependent transverse magnetic fields, respectively. In particular, here we focus on case with $
\vec{B}_{\text{rf}}(t) = B_{\text{rf}} \left[ \cos (\omega t) \hat{x} + \sin (\omega t) \hat{y}\right]  $, where $\omega$ is the rotating frequency of the transverse magnetic field, also known as Radio-Frequency (rf) field. The coupling energy of the system with the field allows us to write the system Hamiltonian as
\begin{equation}
H\left( t\right) = - \vec{\mu}\cdot \vec{B}\left( t\right) =\frac{\hbar\omega _{0}}{2%
}\sigma _{z}+\frac{\hbar\omega _{\text{rf}}}{2}\left[ \cos \left( \omega t\right) \sigma
_{x}+\sin \left( \omega t\right) \sigma _{y}\right] \text{ , }
\label{ApHamiltonian}
\end{equation}
where we used that $\vec{\mu} = \gamma \hbar \vec{\sigma}/2 $ is the magnetic dipole moment, with $\vec{\sigma} = \sigma_{x}\hat{x}+ \sigma_{y}\hat{y} + \sigma_{z}\hat{z}$ and $\gamma$ being the gyromagnetic ratio, which defines the coupling of the spin with the external magnetic field, $\omega _{0}$ is the Larmor frequency and $\omega _{\text{rf}}$ is the coupling of the spin with rf-field. Mathematically we write $\hbar \omega _{0} = -\gamma B_{0}$ and $\hbar \omega _{\text{rf}} = -\gamma B_{\text{rf}}$~\cite{Sarthour:Book}. The procedure discussed in this Appendix is used to find the solution of the spin dynamics when the rf-field is much smaller than the static longitudinal field $ \vec{B}_{0}$, that is, $||\vec{B}_{0}||\gg ||\vec{B}_{\text{rf}}(t)||$ or $|\omega_{0}|\gg |\omega_{\text{rf}}|$. As we shall see in this Appendix, even in case where the rf-field is much smaller than the static field, we will show how the resonance phenomena develops a significant role in spin dynamics of the system. Thus, here we are interested in solving the dynamics
\begin{align}
i\hbar \ket{\dot{\psi}(t)} = H(t)\ket{\psi(t)} \text{ . } \label{ApSchrEq}
\end{align}

To this end, in order to simplify the above equation, let us consider an auxiliary state $\ket{\xi(t)}$ so that $\ket{\psi(t)}$ can be written as
\begin{align}
\ket{\psi(t)} = U_{0}(t)\ket{\xi(t)} \text{ , } \label{ApSol}
\end{align}
where $U_{0}(t)$ is a time-dependent operator $U_{0}(t)$ given by
\begin{align}
U_{0}(t) = \exp \left[-i\frac{\omega }{2} t\sigma _{z}\right] \text{ . }
\end{align}

As we shall see, this particular operator allow us to drastically simplify the dynamics equation of the system. Thus, by using Eq.~\eqref{ApSol} in Eq.~\eqref{ApSchrEq}, we get
\begin{align}
i\hbar \frac{d}{dt} \left[ U_{0}(t)\ket{\xi(t)} \right] &= H(t)U_{0}(t)\ket{\xi(t)} \\
i\hbar \left[\dot{U}_{0}(t)\ket{\xi(t)} + U_{0}(t)\ket{\dot{\xi}(t)} \right] &= H(t)U_{0}(t)\ket{\xi(t)} \\
i\hbar U_{0}(t)\ket{\dot{\xi}(t)}  &= H(t)U_{0}(t)\ket{\xi(t)} - i\hbar \dot{U}_{0}(t)\ket{\xi(t)} \text{ , }
\end{align}
so that we use the property $U^{\dagger}_{0}(t)U_{0}(t)=\1$ to write
\begin{align}
\ket{\dot{\xi}(t)} &= U^{\dagger}_{0}(t)H(t)U_{0}(t)\ket{\xi(t)} - i\hbar U^{\dagger}_{0}(t)\dot{U}_{0}(t)\ket{\xi(t)} \text{ . }
\end{align}

In conclusion, we get a dynamics equation for $\ket{\xi(t)}$ given by
\begin{align}
\ket{\dot{\xi}(t)} &= H_{\xi}(t)\ket{\xi(t)} \text{ . }
\end{align}
in which we define the new (transformed) Hamiltonian
\begin{align}
H_{\xi}(t) = U^{\dagger}_{0}(t)H(t)U_{0}(t) - i\hbar U^{\dagger}_{0}(t)\dot{U}_{0}(t) \text{ . }
\end{align}

It is important to highlight here the additional term in the above equation. The
contribution $i\hbar U^{\dagger}_{0}(t)\dot{U}_{0}(t)$ in $H_{\xi}(t) $ can be interpreted as a ``fictitious potential"~\cite{Klink:97} and it has been considered in a number of applications~\cite{Sarthour:Book,Nielsen:Book,Hu-18-b}. Therefore, we can find the solution $\ket{\psi(t)}$ if we know the solution for $\ket{\xi(t)}$. The most important result here is associated with the Hamiltonian $H_{\xi}(t)$. Now, let us compute the first term in Hamiltonian $H_{\xi}(t)$ as 
\begin{align}
e^{i\frac{\omega }{2}t\sigma _{z}}H\left( t\right) e^{-i\frac{\omega }{2}t\sigma _{z}} &=\frac{\hbar\omega _{0}}{2} e^{i\frac{\omega }{2}t\sigma _{z}}\sigma _{z} e^{-i\frac{\omega }{2}t\sigma _{z}} +\frac{\hbar\omega _{\text{rf}}}{2}\cos \left( \omega t\right)
e^{i\frac{\omega }{2}t\sigma _{z}}\sigma _{x}e^{-i\frac{\omega }{2}t\sigma _{z}} \nonumber \\
&+\frac{\hbar\omega _{\text{rf}}}{2}\sin \left( \omega t\right) e^{i\frac{\omega }{2}t\sigma _{z}}\sigma _{y}e^{-i\frac{\omega }{2%
	}t\sigma _{z}} \nonumber \text{ , }
\end{align}
where we can use the relations
\begin{align}
e^{i\frac{\omega }{2}t\sigma _{z}}\sigma _{x} e^{-i\frac{\omega }{2}t\sigma _{z}} & = \cos \left( \omega t\right) \sigma_{x}-\sin \left( \omega t\right) \sigma _{y} \\
e^{i\frac{\omega }{2}t\sigma _{z}}\sigma _{y} e^{-i\frac{\omega }{2}t\sigma _{z}} & = \cos \left( \omega t\right) \sigma _{y}+\sin \left( \omega t\right)
	\sigma _{x} \text{ , }
\end{align}
to write
\begin{align}
e^{i\frac{\omega }{2}t\sigma _{z}}H\left( t\right) e^{-i\frac{\omega }{2}t\sigma _{z}} &= \hbar \frac{\omega _{0}}{2}\sigma _{z} + \hbar\frac{\omega _{\text{rf}}}{2}\cos \left( \omega
t\right) \left[ \cos \left( \omega t\right) \sigma _{x}-\sin \left( \omega
t\right) \sigma _{y}\right] \\
&+ \hbar\frac{\omega _{\text{rf}}}{2}\sin \left( \omega t\right) \left[ \cos \left(
\omega t\right) \sigma _{y}+\sin \left( \omega t\right) \sigma _{x}\right] \nonumber \\
&= \hbar \frac{\omega _{0}}{2}\sigma _{z}+ \hbar \frac{\omega _{\text{rf}}}{2}\sigma _{x} \text{ . }
\end{align}

Now, by computing the second term of $H_{\xi}(t)$ we get
\begin{align}
i\hbar U^{\dagger}_{0}(t)\dot{U}_{0}(t) = i\hbar e^{i\frac{\omega }{2}t\sigma _{z}} \frac{d}{dt}\left[  e^{-i\frac{\omega }{2}t\sigma _{z}}\right] = i\hbar e^{i\frac{\omega }{2}t\sigma _{z}} \left[ -i\frac{\omega }{2}\sigma _{z} e^{-i\frac{\omega }{2}t\sigma _{z}}\right] = \hbar \frac{\omega }{2}\sigma _{z} \text{ . }
\end{align}

Thus, we conclude that 
\begin{align}
H_{\xi}(t) &= \tilde{H} = \hbar \frac{\omega _{0}-\omega }{2}\sigma_{z}+ \hbar \frac{\omega _{\text{rf}}}{2}\sigma _{x} \text{ . }
\end{align}

It is worth to highlight in above equation the quantity $\Delta = \omega_{0} - \omega$, called detuning between the rf-field and the Larmor frequencies. In particular, when $\Delta$ is very close to zero we have a \textit{resonance situation}, and this configuration is very important because if we put $\Delta \rightarrow 0$, the dominant parameters in dynamics of the system are associated with the rf-field. Moreover, this result is very useful, since we now can compute the solution for $\ket{\xi(t)}$ from a time-\textit{independent} Hamiltonian. Now, the solution for $\ket{\xi(t)}$ reads as $\ket{\xi(t)} = U_{xz}(t)\ket{\xi(0)}$, where
\begin{align}
U_{xz}(t) &= \exp \left[ -\frac{i}{\hbar }\tilde{H}t\right] = \begin{bmatrix}
\cos \left(\frac{\Omega t}{2}\right) + i\frac{\Delta}{\Omega}\sin \left(\frac{\Omega t}{2}\right) & i\frac{\omega_{\text{rf}}}{\Omega}\sin \left(\frac{\Omega t}{2}\right)\\
i\frac{\omega_{\text{rf}}}{\Omega}\sin \left(\frac{\Omega t}{2}\right) & \cos \left(\frac{\Omega t}{2}\right) - i\frac{\Delta}{\Omega}\sin \left(\frac{\Omega t}{2}\right)
\end{bmatrix} \text{ , }
\end{align}
where $\Omega^2 = \Delta^2+\omega_{\text{rf}}^2$ is the effective Rabi frequency, which takes into account effects due to the detuning $\Delta$. Thus, by using the fact that $U_{xz}(0) = U_{0}(0) = \1$, we find that $\ket{\xi(0)} = \ket{\psi(0)}$ and this calculation concludes the proof of the Eq.~\eqref{SingSpinDyn}.

	\section{Non-unitary effects in superconducting transmon qubits} \label{NonUnitaryAp}
	
	In this section we briefly discuss about undesired effects in real quantum information processors. In particular, we will focus on experimental device used in this manuscript.
	
	\subsection{Decoherence phenomena} \label{DecoherenceAp}
	
	Due to our inability of decoupling our quantum system from its environment (thermal reservoir, for example), the run time of a computation (or dynamics) is bounded by two important time scales, namely, $T_{1}$ and $T_{2}$. The first quantity $T_{1}$ is an energy relaxation time scale that arises from natural perturbations (external effects) in the system. This natural perturbation drives the system through a non-unitary process from an arbitrary state $\ket{\psi}$ to the fundamental state $\ketds$. This process is also called \textit{spontaneous emission}~\cite{Koch:07}. Therefore, since quantum computation needs coherence (superposition) to implement powerful algorithms, we can see that such effect becomes a problem. As an alternative to bypass this problem, we can implement computations with run time $T_{\text{comp}} \ll T_{1}$, so that the decohering effect due to spontaneous emission is minimized (or suppressed in some cases). On the other hand, there is a different time scale $T_{2}$ that is associated with loss of coherence of the system (also called \textit{phase damping}~\cite{Nielsen:Book}). We mean, given a general quantum state $\ket{\psi} = a\ketds + b\ketus$, the coherence of the system in the basis $\{\ketds,\ketus\}$ can be easily identified from the density matrix
	\begin{align}
	\rho_{\psi} = \ket{\psi}\bra{\psi} = \begin{bmatrix}
	|b|^2 & b a^{\ast} \\
	a b^{\ast} & |a|^2\\
	\end{bmatrix} \text{ , }
	\end{align}
	where the diagonal terms ($|b|^2$ and $|a|^2$) denote the population of each state $\{\ketds,\ketus\}$ and off-diagonal state are \textit{coherence terms}, that is, we say that a quantum state has coherence in basis $\{\ketds,\ketus\}$ when $|a b^{\ast}| \neq 0$. Therefore, non-unitary processes that drastically change the quantity $|a b^{\ast}|$ are undesired. In order to explain the phase damping process, consider the initial state of the system is given by the above density matrix, then if the system evolves under phase damping process the evolved state $\rho(t)$ of the system state is~\cite{Nielsen:Book}
	\begin{align}
	\rho(t) = \begin{bmatrix}
	|b|^2 & a b^{\ast} e^{-t/2T_{2}} \\
	ba^{\ast} e^{-t/2T_{2}} & |a|^2\\
	\end{bmatrix} \text{ . }
	\end{align}
	
	Therefore, from above equation we can see that $\rho(t \gg T_{2}) \approx \text{diag}\begin{bmatrix}
	|b|^2 & |a|^2 \end{bmatrix}$, where the system becomes (approximately) a \textit{maximally mixed state} and no coherence in basis $\{\ketds,\ketus\}$ is present. On the other hand, we can see that $\rho(t \ll 2T_{2}) \approx \rho_{\psi}$, so that the loss of coherence can be negligible. For this reason, the knowledge of the quantities $T_{1}$ and $T_{2}$ is very important to the run time of the computation $T_{\text{comp}}$. The bigger the $T_{1}$ and $T_{2}$ are, the bigger the $T_{\text{comp}}$ can be. In the experiment presented in this manuscript, the computation time is of order of a few tens of nanoseconds, while the time scales $T_{1}$ and $T_{2}$ are of the order of a few of tens of microseconds. Then, the condition $T_{\text{comp}} \ll T_{1}$ and $T_{\text{comp}} \ll T_{2}$ is verified and we can approximate our analysis via unitary processes, as done in main text of the manuscript. Consequently, any error in process is associated with systematic control and readout errors, as we shall discuss in next section.
	
	\subsection{Control and readout of transmon superconducting qubits} \label{ErrorAp}
	
	Any physical system (two-level system) used as a qubit for quantum information processing needs to satisfy a number of criteria known as \textit{DiVincenzo's criteria}~\cite{DiVincenzoCriteria}. DiVincenzo's criteria is a list with 7 conditions to be satisfied for any system that can be used to quantum information, computation and communication. In particular, two of these conditions are related with control and readout of the qubits of the system, that is, any good qubit needs to be manipulated and the information encoded in it needs to be read. In particular, in transmon superconducting qubit the control and the readout can be achieved by an external driving microwave acting on the system. For more details about the interaction of the transmon qubit with the external driving field, we recommend the pioneer manuscript~\cite{Koch:07}. 
	
	The qubit operations (gates) are implemented by means of microwave pulses at the qubit transition frequency, where the microwave field couples to the the qubit via an electromagnetic coupling. In this situation, the qubit state is driven through Rabi oscillations. Therefore, imprecise control of these pulses (phase or amplitude, for example) reflects on the optimal control of the qubit operations, hence some error gate is evidently possible in this case. In the device considered in this manuscript, the error gate is of the order of~$\sim10^{-3}$, that is, the gate is efficiently implemented with fidelity $F \sim 0.999$ ($99.9\%$). More specifically, the fidelity of the gate applied to the qubit used in our first experiment (Table~\ref{TableResonance}) is $F = 0.99536$ ($99.536\%$), since the error gate is $0.00464$, and in second experiment we have $F = 0.99579$ ($99.579\%$), because the error gate is $0.00421$ (Table~\ref{TablePrecession}).
	
	On the other hand, the readout is obtained by measuring the phase or amplitude of the transmitted radiation of a microwave drive field~\cite{Koch:07}. To do that, a readout microwave interacts with qubit via a dispersive interaction. Then, the measurement of the effective transition frequency of the resonator is used to determinate the qubit state. It is important to mention that, since the measurement is effectively done on the resonator, this process is called \textit{quantum nondemolition measurement}~\cite{wallraff2004}. Due to the complexity of the measurement process, such process is more sensitive to errors then the qubit operations previously discussed. To the qubit used in our experiment, we have the error given by $0.017$ and $0.016$ for the first and second experiment, respectively. This means that the fidelity readout is $98,3\%$ and $98,4\%$, respectively.

\end{document}